\documentclass[twocolumn,prl,superscriptaddress,notitlepage]{revtex4-1}

\usepackage{graphicx} 
\usepackage{booktabs}
\usepackage{siunitx}
\usepackage{url}
\usepackage{adjustbox}
\usepackage{float}
\usepackage{amsthm}
\usepackage{mathtools}
\usepackage{physics}
\usepackage{xcolor}
\usepackage{graphicx}
\usepackage{multirow}

\usepackage{adjustbox}
\usepackage{placeins}
\usepackage[T1]{fontenc}
\usepackage{lipsum}
\usepackage{csquotes}
\usepackage{booktabs}
\usepackage{siunitx}

\newcolumntype{d}{S[
    input-open-uncertainty=,
    input-close-uncertainty=,
    parse-numbers = false,
    table-align-text-pre=false,
    table-align-text-post=false
 ]}

\begin{document}
\title{The diminishing state of shared reality \\ on US television news}

\author{Homa Hosseinmardi}
\affiliation{University of Pennsylvania, Philadelphia, PA 19104}
\author{Samuel Wolken} 
\affiliation{University of Pennsylvania, Philadelphia, PA 19104}
\author{David M. Rothschild}
\affiliation{Microsoft Research New York, New York, NY 10012}
\author{Duncan J. Watts}
\affiliation{University of Pennsylvania, Philadelphia, PA 19104}


\begin{abstract}
The potential for a large, diverse population to coexist peacefully is thought to depend on the existence of a ``shared reality:'' a public sphere in which participants are exposed to similar facts about similar topics. A generation ago, broadcast television news was widely considered to serve this function; however, since the rise of cable news in the 1990s, critics and scholars have worried that the corresponding fragmentation and segregation of audiences along partisan lines has caused this shared reality to be lost. Here we examine this concern using a unique combination of data sets tracking the production (since 2012) and consumption (since 2016) of television news content on the three largest cable and broadcast networks respectively. With regard to production, we find strong evidence for the ``loss of shared reality hypothesis:'' while broadcast continues to cover similar topics with similar language, cable news networks have become increasingly distinct, both from broadcast news and each other, diverging both in terms of content and language. With regard to consumption, we find more mixed evidence: while broadcast news has indeed declined in popularity, it remains the dominant source of news for roughly 50\% more Americans than does cable; moreover, its decline, while somewhat attributable to cable, appears driven more by a shift away from news consumption altogether than a growth in cable consumption. We conclude that shared reality on US television news is indeed diminishing, but is more robust than previously thought and is declining for somewhat different reasons. 
\end{abstract}

\maketitle

As recently as the 1970s, an estimated 27-29 million people (out of a population of around 200 million) tuned in on any given weekday to watch Walter Cronkite deliver the news on CBS~\cite{decline_network}. Along with his peers ABC and NBC---which together comprised the ``big three'' broadcast news---Cronkite's dominance created a kind of ``shared reality'' for a substantial portion of the politically informed and active US population. 
Even if they disagreed about the meaning of what they heard or what they should do about it, for 30 minutes every evening Americans across the political spectrum consumed the same set of facts about the same set of events, many of them from the same trusted individual. 
In 2018, former President Barack Obama claimed that this shared reality was gone, describing partisan media consumers as ``living on another planet'' and concluding that ``one of the biggest challenges we have to our democracy is the degree to which we don’t share a common baseline of facts'' \cite{Chandran_2018}. 
Obama's comment echoed broadly shared concerns among scholars of media and democracy. News media influences viewers by shaping how they assess the importance of political issues \cite{mccombs_shaw_1972,iyengar_kinder_1987, zaller92}, formulate policy opinions \cite{broockman2022impacts}, talk about politics \cite{king17}, evaluate politicians' performance \cite{krosnick_kinder_1990}, vote \cite{martin17}, or decide whether or not to take precautionary measures against the spread of COVID-19 \cite{kim2020}. 
Thus, when viewers of different partisan leanings consume their news from distinct sources, and when those sources offer incommensurate views of the world, two related concerns arise: first, that citizens will be exposed to less diverse information and hence be less informed overall; and second that opposing ``sides'' will fracture along many dimensions simultaneously, potentially corroding support for democratic norms among voters \cite{finkel2020} and a taste for compromise among governing elites \cite{jamieson2010}.  

To date, empirical evidence regarding the seriousness of these concerns has been inconclusive. Although many scholars have worried about the fragmentation of the mass public into ``echo chambers'' or ``filter bubbles'' in online environments~\cite{pariser2011, sunstein2017}, systematic empirical analyses have found that  Americans' online news consumption, while surprisingly low~\cite{allen2020evaluating}, is generally diverse \cite{eady2019, guess2020, guess2021, guess2018, francisci2021,bakshyExposureIdeologicallyDiverse2015,flaxman2016filter}, and that at most a tiny percentage of people plausibly ``live'' in echo chambers~\cite{muise2022quantifying}. Media fragmentation in television is arguably a bigger concern, in part because Americans spend, on average, about five times as much time consuming news content through television than on the internet \cite{allen2020evaluating}, and in part because almost ten times as many Americans have partisan-segregated news diets in their television news consumption than in their online news consumption \cite{muise2022quantifying}. 
Recent research also suggests that bias in television news production has become more pronounced in recent years \cite{kim_lelkes_mccrain} and that such biases could causally impact viewers' attitudes \cite{dellavigna07,broockman2022impacts, martin17, martin2019}. Most of the research on television has focused on consumption of cable news, with a few notable papers exploring the production of cable news~\cite{kim_lelkes_mccrain,broockman2022impacts}, and very little on broadcast news in either capacity. We note, however, that while these studies suggest that segregation of television news audiences may indeed be problematic they do not directly address the issue of the putative disappearance of shared reality. 

In this paper, we build on this prior work by defining shared reality as it pertains to US television news and studying its evolution over recent history. Our definition of shared reality seeks to quantify the extent to which Americans consume similar information about similar topics, and hence it incorporates two related components: one regarding the production of media; and the other regarding its consumption. On the production side, by shared reality we mean (a) that stations exhibit similar ``selection bias,'' meaning that allocate similar levels of attention to potential topics, and (b) that conditional on covering the same topic, they exhibit similar ``framing bias,'' meaning that they use similar language to talk about it. By contrast, if different stations selectively cover different topics, or if when they do cover the same topics, they frame them differently via the language they choose, then we would define that as an absence of shared reality. 
On the consumption side, meanwhile, what we mean by shared reality implies either that different people consume their news from the same source, or that they consume it from different sources that are largely interchangeable in terms of their selection and framing of topics. By contrast, if fewer Americans than in the past consume media from common sources or if they have gravitated from largely interchangeable sources to increasingly polarized sources, then we would call that a loss of shared reality. 

Reflecting the two-sided nature of the shared reality problem, our analysis leverages two unique data sets: one representing production of news and the other representing its consumption. Our production data comes from a comprehensive corpus of television program transcripts covering all types of programming and spanning all 210 designated market areas (DMAs) across the US. From this data, we can extract the closed caption text of every unique episode ($328,432$) of every national news program ($5,889$) broadcast by any of 
national news content on each of the three broadcast networks (ABC, CBS, and NBC) and the three primary cable news channels (CNN, Fox News Channel [FNC], and MSNBC)
over the course of almost a decade (December 2012 to October 2022). For each program, our data also include the program title, date and time of broadcast, and the duration of the program. Further, we categorized all news programs into one of the following categories: hard news, talk shows, partisan/opinion news, soft news, local news, other. Because most news programs last between 30 minutes and one hour, and so tend to cover multiple topics, we further partition each episode into segments of roughly 150 words each (see Materials and Methods for details), yielding $13,446,736$ segments that form the basis of our news production analysis. 

Our consumption data comes from a large (average $N\approx 114,977$ per month) US representative panel that details individual-level television consumption, both for viewers and programs, spanning almost seven years from January 2016 to October 2022. From this viewing data, we can directly measure the number of minutes per day each panel member watched national news content on 
the same six stations, as well as their total daily minutes spent watching television news (including local news, other cable news channels, etc.) and television in general (including non-news content such as sports, reality shows, etc.). We note that these six stations account for the overwhelming majority of American television news consumers: in 2022, 90\% of US adults who watched at least 30 minutes of television news in a given month consumed content from broadcast or cable news networks---a share that increased during the period of our data (in 2016 it was 85\%). Thus, our data, which tracks an average of about 9 million hours of television viewing per month (roughly three hours per panelist per day), includes a reasonably complete picture of US television news consumption. 

In the remainder of this paper, we analyze over time trends in both the production and consumption sides of the market for television news. 
On the production side, we study two related questions. First, we study \emph{how much} each station talks about each of twenty four politically relevant topics, finding that broadcast stations allocate their attention in relatively similar way whereas cable stations have increasing diverged along partisan lines among themselves as well as collectively from broadcast. Second, we study \emph{how distinctive} a station's coverage of any given topic is relative to other stations, finding again that broadcast is relatively interchangeable whereas cable is again increasingly distinct and polarized along partisan lines. On the consumption side, we then study the evolution of the US audience for news and how its attention has shifted over time, finding that the audience for broadcast has eroded substantially but remains the dominant source of news for Americans. Moreover, the decline in broadcast appears less attributable to migration to cable than to an overall shift away from news consumption of any kind. Finally, we conclude with a brief discussion of the implications of our findings and possible directions for future work.



\begin{figure*}[tb!]
\centering
\includegraphics[width=16cm]{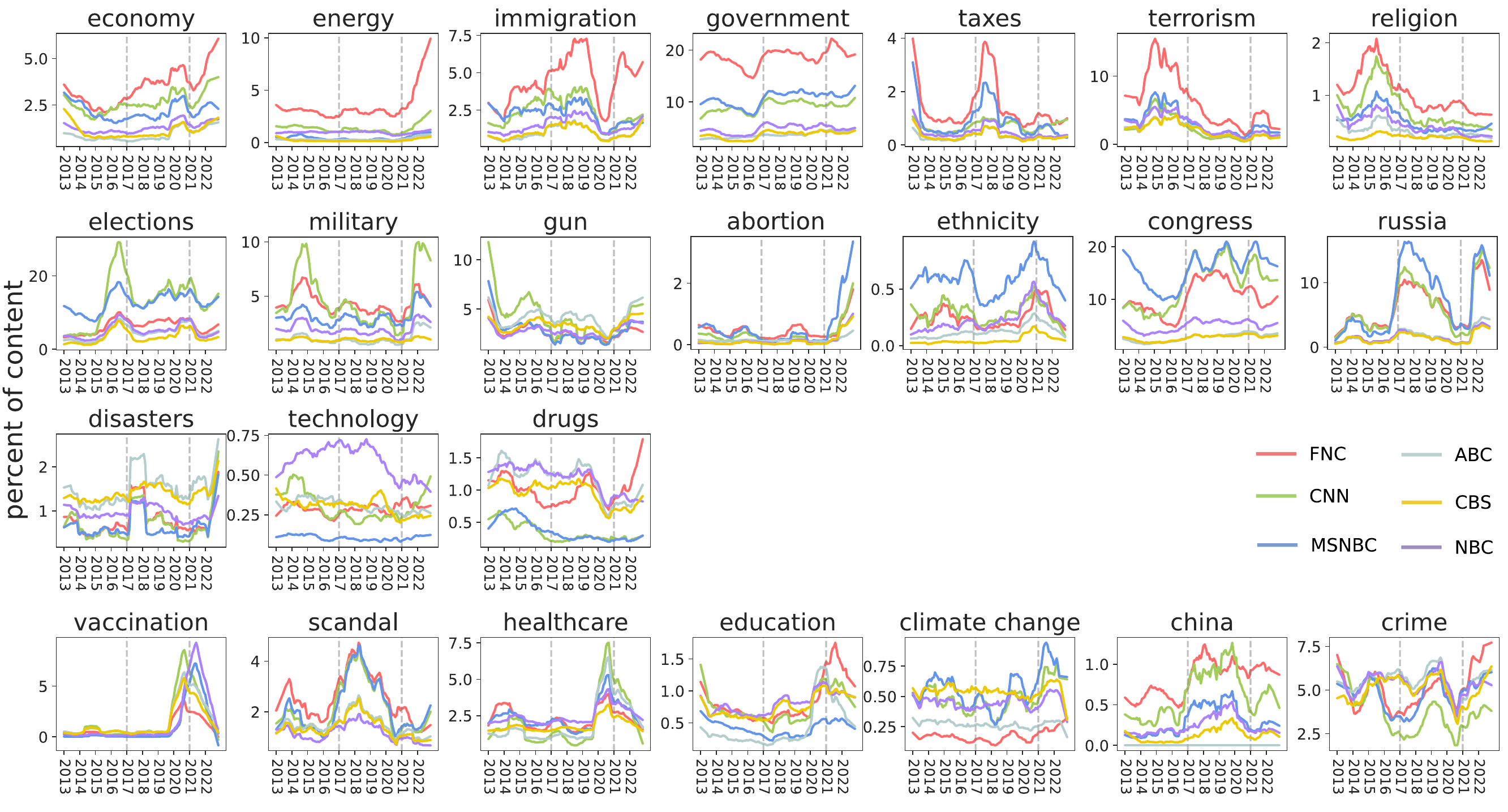}
\vspace{-3mm}
\caption{Topics as a percentage of channel's news content for all 24 topics.}
\label{fig:topics_selection}
\end{figure*}

\section*{Topic selection across stations}

To estimate the presence and severity of topic selection across different stations, we computed the proportion of news airtime devoted by each station to 24 topics that cover a combination of socially polarizing issues (e.g. abortion, immigration, gun, climate change), issues that became salient due to specific events (e.g. vaccination, China, Russia), and issues that are of perennial relevance to US politics (e.g. healthcare, energy, education, the economy) and society (e.g. disasters, drugs, crime). 
To obtain these estimates, we first coded all $13,446,736$ news segments using a novel two-layered human-in-the-loop classification model that achieved an average precision of 78.45\%. We then estimated the proportion of coverage that is about a certain topic, $z$, on a given day by dividing the total number of words of all segments labeled as topic $z$ by the total number of words in that station's news coverage on the same day (see Materials and Methods for more information about the classification process). Together, the 24 topics in Fig. \ref{fig:topics_selection} account for 55.1\% of news segments from cable and 27.6\% of news segments from broadcast networks. 

Fig. \ref{fig:topics_selection} shows the proportion of airtime devoted by the six stations to each of the 24 topics, where the top row shows topics dominated by FNC (red), the second row by CNN (green) and MSNBC (blue), and the third row by broadcast news (ABC, grey; CBS, yellow; and NBC, purple). The fifth row shows topics that did not consistently receive more coverage by one source than the others. Thus, FNC consistently paid more attention to the economy, energy, immigration, government, taxes, terrorism, and religion, while CNN and MSNBC paid more attention to elections, the military, guns, abortion, ethnicity, congress, and Russia. Meanwhile, the broadcast networks paid more attention to disasters, technology, and drugs. Finally, we note that while no station paid obviously more attention to vaccination, scandals, healthcare, education, climate change, China, or crime, FNC paid distinctly less attention to climate change than any other station. 

In some cases, the differences observed in Fig. \ref{fig:topics_selection} appear to reflect ``partisan selective filtering''~\cite{broockman2022impacts}, wherein partisan cable channels preferentially cover issues that favor their viewers' preferred political party~\cite{krosnick_kinder_1990}. For example, FNC's disproportionate attention to the economy, taxes, immigration, and terrorism aligns with Republican politicians' incentives. Similarly, MSNBC's focus on abortion and ethnicity (which may evoke racially coded issues such as welfare and poverty) boosts political topics that benefit Democratic politicians. Partisan topic selection is especially apparent for salient political topics, with cable networks devoting a large share of their coverage to hot-button issues that resonate with their audience's political predispositions. The Supreme Court decision to overturn Roe v. Wade led to a spike in coverage of abortion across stations, but MSNBC dedicated a far greater share of coverage to the topic. FNC, meanwhile, featured more coverage of immigration than the other stations, especially during the Biden presidency, in which they devoted approximately three times more airtime to immigration than the next closest station. The Russia investigation dominated airwaves for more than a year of Trump's presidency, but no station's surge in Russia coverage matched that of MSNBC. On the contrary, the broadcast networks tend to lead in coverage of issues with less overt partisan coding: disasters, technology, and drugs.

Surges in coverage of specific topics often map onto high-profile news events. For example the ``Russia'' topic tracks the Russian intervention in the 2016 election and subsequent investigation and then peaks again when Russia invaded Ukraine. ``Vaccination'' tracks with the introduction of the COVID-19 vaccine.  ``Taxes'' pops around the time of the 2017 Trump tax cuts. Still, substantial variation exists across stations even in these event-driven topics. While every station is lock-step in covering vaccinations, the cable networks devoted more of their news coverage to the Trump-Russia investigation, for example, than did the broadcast networks. More generally, the share of coverage that stations devoted to other topics, such as ``climate change'', ``crime'', and ``drugs,'' are not tied to particularly distinct news events, allowing even greater variation in the coverage by the various stations.


\begin{figure}[tb!]
\centering
\includegraphics[width=8cm]{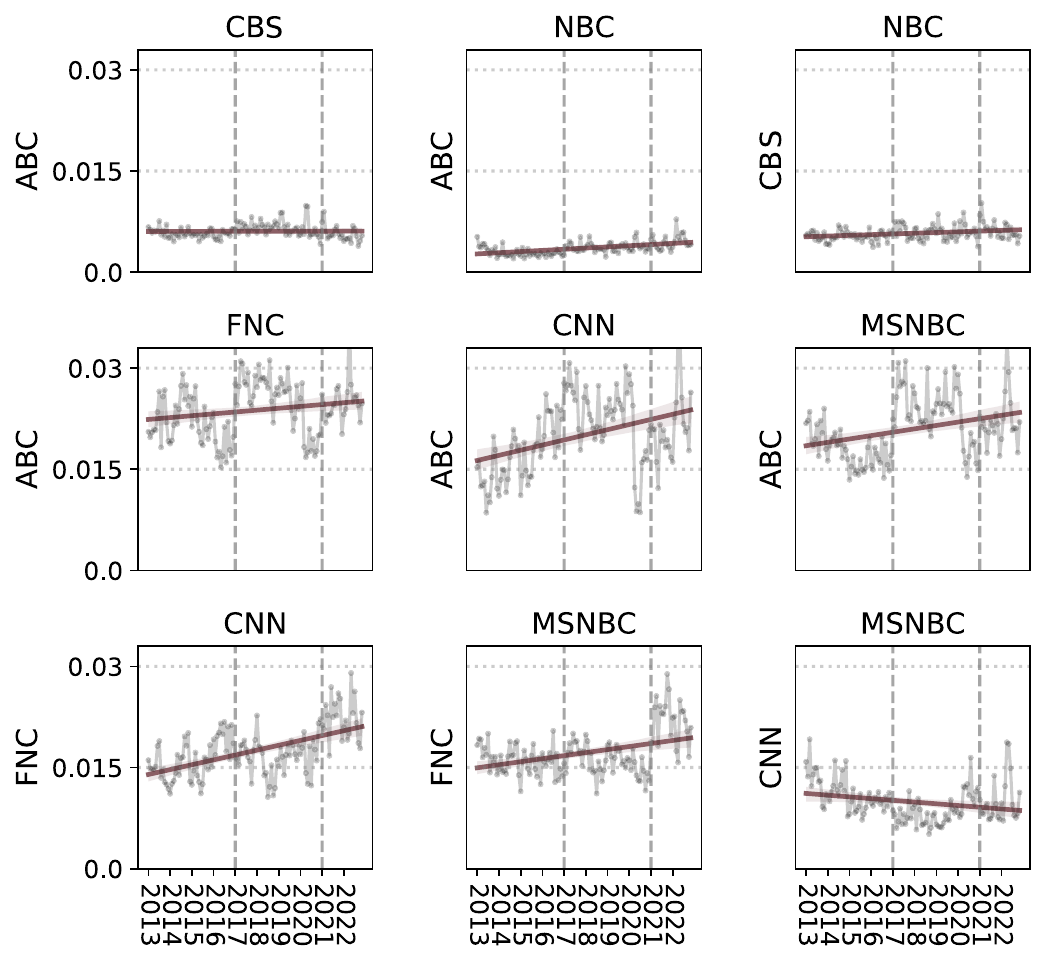}
\vspace{-2mm}
\caption{Average differences in coverage (percent of time) for the 24 topics between all six stations.}
\label{fig:aggregate_topics_selection}
\end{figure}

Fig.~\ref{fig:topics_selection} clearly shows differences in emphasis between stations at the level of individual topics; however, it does not clearly show the degree of aggregate variation in topic selection---i.e. between-station differences aggregated over all 24 topics---or whether such variation increased over time. 
To address this issue, for each pair of stations $i$ and $j$ and at each time $t$, we computed the aggregate difference in topic selection, $\delta^{(i,j)} (t) = \sum_{k=1}^{K} |y^{i}_k(t)-y^{j}_k(t)| / K$,  where $y^{i}_{k(t)}$ is the proportion of time devoted by station $i$ to $k^{\text{th}}$ topic  at time $t$, and $K=24$ is total number of topics. Fig.~\ref{fig:aggregate_topics_selection} shows this aggregated measure of difference in topic selection over time, where the vertical axis tracks the magnitude of difference. 
The top row of Fig.~\ref{fig:aggregate_topics_selection} shows differences in topic selection among the broadcast news programs: CBS vs. ABC (left); NBC vs. ABC (middle); and NBC vs. CBS (right). In all cases, the pairwise difference is small and not increasing over the time period, consistent with the visual impression from Fig.~\ref{fig:topics_selection} that the broadcast networks devote similar amounts of coverage to a wide range of topics.
Next, the second row of Fig.~\ref{fig:aggregate_topics_selection} compares each of the three cable networks to ABC national news programming. In contrast with the between-broadcast comparisons in the top row, there are large differences in topic selection between broadcast news and all three cable networks---differences that are increasing over time. FNC is the most distinct overall, both starting and ending the ten-year period with higher values of $\delta^{(i,j)}$ than either CNN or MSNBC. Perhaps surprisingly, both MSNBC and especially CNN show steeper increases over the interval: in 2012, MSNBC and CNN had more similar topic selection to ABC than FNC did; however, by 2022, all three cable networks are similarly distinct from brodcast news. 
Finally, the bottom row of Fig~\ref{fig:aggregate_topics_selection} shows pairwise differences in topic selection between the cable networks: CNN vs. FNC (left); MSNBS vs. FNC (middle); and MSNBC vs. CNN. While the differences between the cable networks are not as large as the difference between cable and the broadcast networks, as expected, both CNN and MSNBC display substantial differences with FNC where these differences have grown with time. In contrast, CNN and MSNBC were relatively similar in 2012 and have grown more similar since then. 
 
 Together, Figs.~\ref{fig:topics_selection} and~\ref{fig:aggregate_topics_selection} show that viewers of broadcast news received largely interchangeable news regardless of which station they watched or when in time they watched it. In contrast, viewers of cable news saw a different mix of topics than those watching the broadcast networks, where these differences increased over time. FNC viewers saw more coverage of topics emphasized by Republican politicians (such as immigration and the economy), whereas CNN and MSNBC viewers watched news coverage that converged on topics from Democratic politicians' agenda (such as abortion, ethnicity, and the Russia investigation). By 2022, FNC came to stand on its own as distinct both from the broadcast networks and also the other two cable channels in how it allocates attention to topics.


\begin{figure}[tb!]
\centering
\includegraphics[width=8.5cm]
{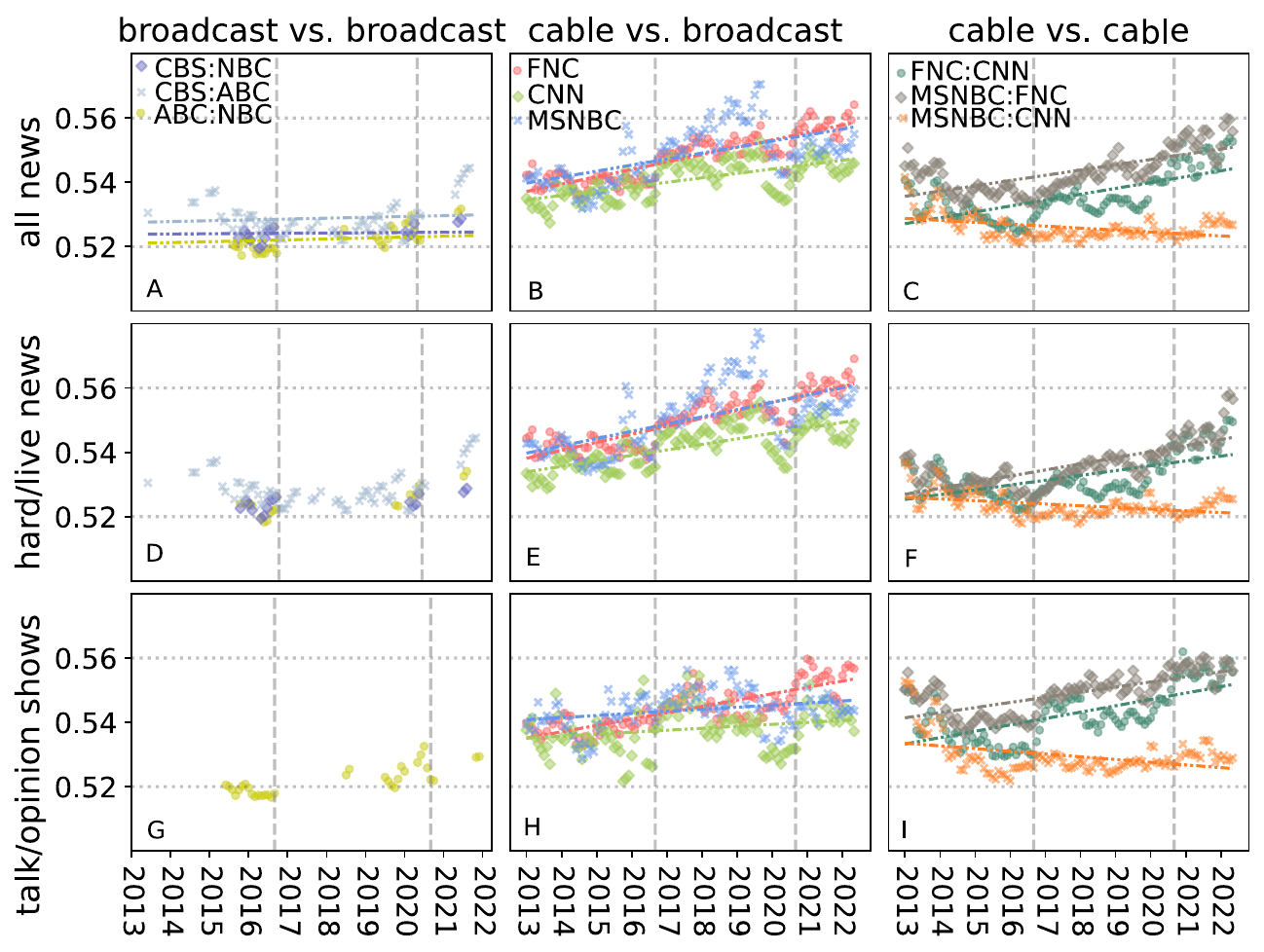}
\vspace{1mm}
\caption{Averaged within-topic: (i) polarization between NBC vs. ABC (yellow), NBC vs. CBS (purple), and CBS vs. ABC (blue) over time for all news (A), hard news shows (D), and talk/opinion shows  (G);  (ii) polarization of the cable networks—FNC (red), CNN (green), and MSNBC (blue)—with respect to broadcast news over time for all news (B), hard news (E) and talk/opinion shows (H); (iii) within-cable polarization, comparing FNC with CNN (green) and MSNBC (grey), and CNN with MSNBC (orange) for all news (C), hard news (F) and talk/opinion shows (I). }
\label{fig:polarization_over_time}
\end{figure}


\begin{figure*}[tb!]
\centering
\vspace{1mm}
\includegraphics[width=17.5cm]{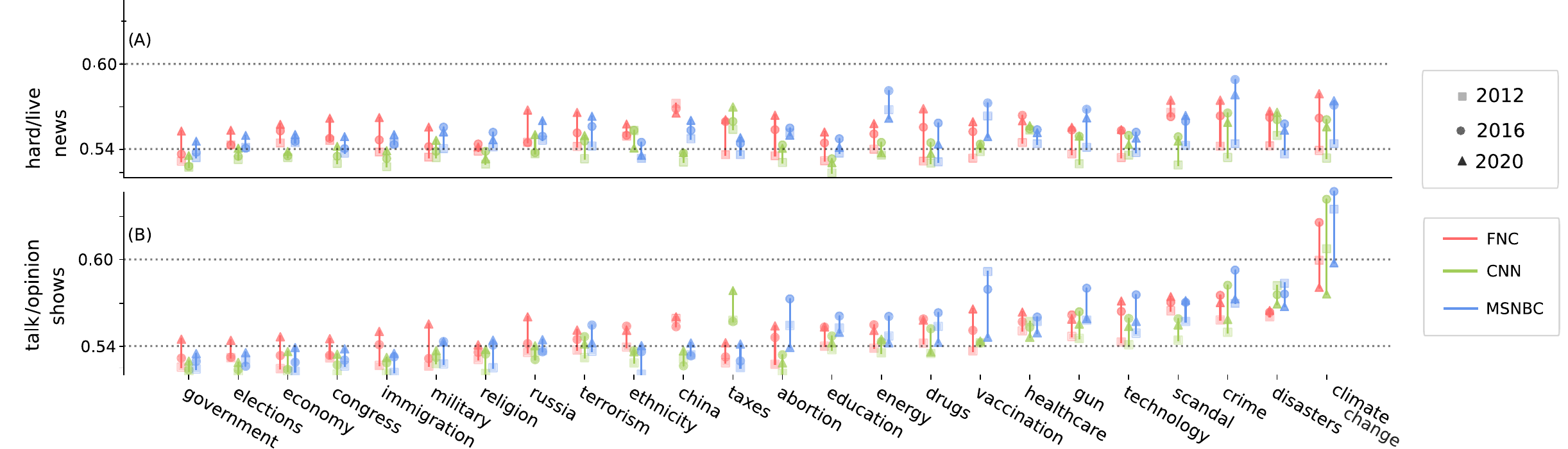}
\caption{Topic-level polarization for FNC, CNN, and MSNBC for talk/opinion shows (A) and hard/live news shows (B) compared with broadcast news networks. }
\label{fig:polarization_top_level}
\end{figure*}

\section*{Polarization in topic coverage}
Although significant, the increasing discrepancy in topic focus potentially understates between-station differences in coverage. Even when two stations cover the same topic they have considerable leeway to talk about it differently, emphasizing different facts, actors, or narratives, and expressing more or less positive vs. negative affect. For example, when discussing immigration, CNN and MSNBC may emphasize the humanitarian issues associated with overcrowded camps of asylum seekers (e.g.,\textit{``we should not be locking these people up in the way were doing it. craig, its atrocious. i think it goes against our values''} from MSNBC), whereas FNC may emphasize border security or criminality (e.g., \textit{``the border patrol apprehending thousands of migrants a day coming across all kinds of criminal activity as well as felons posing a threat to our safety''}). 
Likewise, one station's coverage of vaccinations may celebrate their efficacy while another may rail against perceived safety concerns or social pressures to get vaccinated. In other words, by invoking different language to talk about the same topic, different stations can frame the topic in different ways.

To operationalize differences in the language used by to talk about a given topic by, we adopt a measure proposed by Gentzkow et al.~\cite{gentzkow2019measuring} to quantify the partisanship of congressional speech. 
In that context, the measure $\pi(x)$ corresponds to the average probability that a listener with no ex-ante information can correctly identify their political party (Democrat or Republican) after listening to a single two-word utterance (see Materials and Methods for details). 
Thus, a value of 0.50 implies no discernible difference between the language used by the two parties while higher values correspond to increasingly obvious differences. Historically, Gentzkow et al. found that their measure of partisanship fluctuated between 0.50 and 0.54 from 1870-1990 and then ``exploded'' in the 1990s and 2000s, reflecting a sharp rise in elite polarization that began with the Republican Party's 1994 ``Contract with America,'' and reached a maximum of 0.54 in 2008, the year in which Barack Obama was elected president. In other words, the numerical difference between 0.52 and 0.54 for a single utterance is substantively significant, corresponding to posterior probabilities of 0.54 and 0.73 respectively when the listener was exposed to as little as one minute of speech~\cite{gentzkow2019measuring}. By analogy with Gentzkow et al., we define ``polarization'' $\pi(x)$ in television news as the average probability that a listener with no ex-ante information could correctly identify a given station after listening to a single one-word or two-word utterance.


\begin{figure*}[tb!]
\centering
\includegraphics[width=15.25cm]{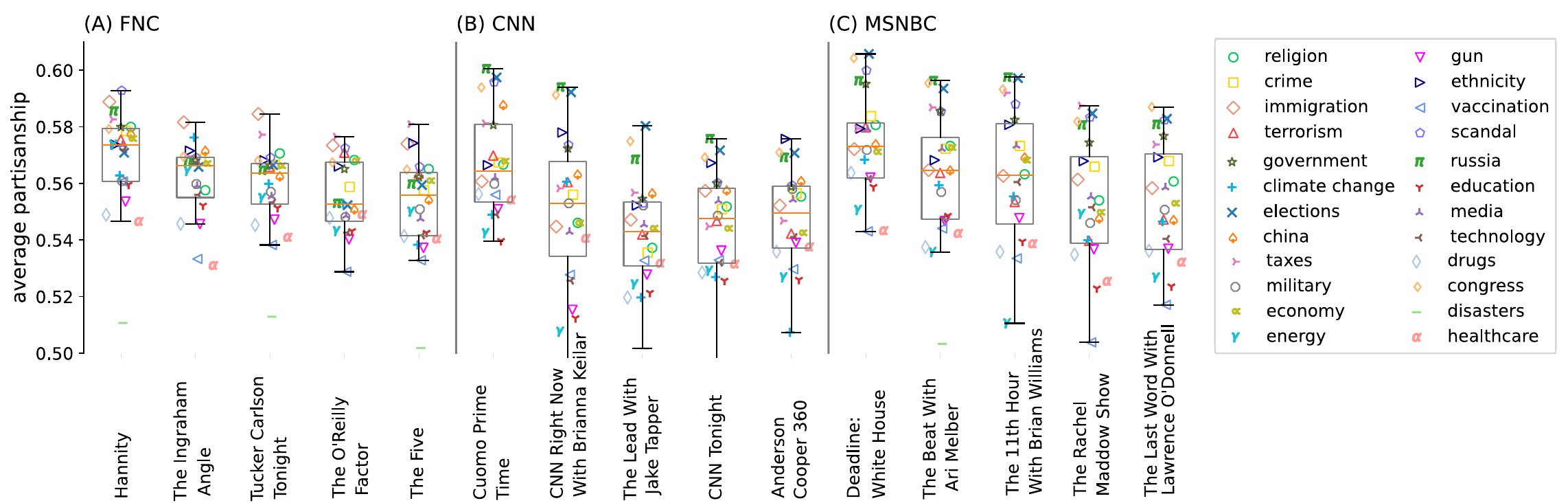}
\vspace{2mm}
\caption{Boxplot of partisan scores at the program level. Markers show the average partisan score of each topic within that specific program.}
\label{fig:program_polarization}
\end{figure*}

Fig.~\ref{fig:polarization_over_time} shows trends in polarization over time within and between broadcast and cable news networks for all news types (panels A-C), hard news only (panels D-F), and talk/opinion shows (panels G-I), which includes shows such as NBC's Morning Joe and CBS's The McLaughlin Group. Referring initially to all news types, Fig.~\ref{fig:polarization_over_time} (A-C) shows three main trends. 
First, Fig.~\ref{fig:polarization_over_time}A, which compares the broadcast networks with one another (NBC vs. ABC (yellow), NBC vs. CBS (purple), and CBS vs. ABC (blue)), shows that, similar to topic coverage (Fig.~\ref{fig:aggregate_topics_selection}), polarization between broadcast networks is low and not increasing discernibly over time.
Second, Fig.~\ref{fig:polarization_over_time}B shows the polarization of the cable networks---FNC (red), CNN (green), and MSNBC (blue)---with respect to broadcast news, where to create our ``broadcast news'' baseline we first combined all utterances from NBC, CBS, and ABC (the results are not substantively different when we substitute with a single broadcast network).  Here we see a starkly different picture: polarization starts higher than the within-broadcast average and then increases by 2.85\% over the time period, where we note that an increase of 0.16 is comparable to the increase from 0.52 to 0.54 that Gentzkow et al. characterized as ``explosive.'' 
High-polarization segments often reflect partisan talking points, for example, ``...\textit{There is no Trump-Russia collusion but Hillary did pay for Russian lies that was used to disseminate lies to the American people}...,'' (0.65) and ``...\textit{Donald Trump went bust. He could not borrow money from banks. It was coming from Russia, and once he's out of office you're going to see money laundering, you're going to see a criminal enterprise}...,'' (0.60) reflect different frames on FNC and MSNBC, respectively. In contrast, low-polarization segments often employ similar language to describe ongoing events: for example, ``...\textit{In 30 minutes, the deputy attorney general and FBI director will testify about 2016 Russian election hacking. }...'' on MSNBC (0.50), and ``...\textit{FBI director and the NSA chief heading back to capitol hill next week. to testify before the house intelligence committee. part of the intelligence into Russian election meddling}'' on FNC (0.48) talk similarly about topic ``Russia.''
Third, Fig.~\ref{fig:polarization_over_time}C shows within-cable polarization, comparing FNC with CNN (green) and MSNBC (grey). Interestingly, at the the beginning of the time period (in 2013) polarization between FNC and the two left-leaning networks is not perceptibly different than it is between CNN and MSNBC themselves, and the overall level is comparable with the broadcast networks. By 2023, however, FNC has diverged strikingly from the other two, while MSNBC and CNN have become less polarized relative to one another.  


\begin{figure*}[tb!]
\centering

\includegraphics[width=16cm]{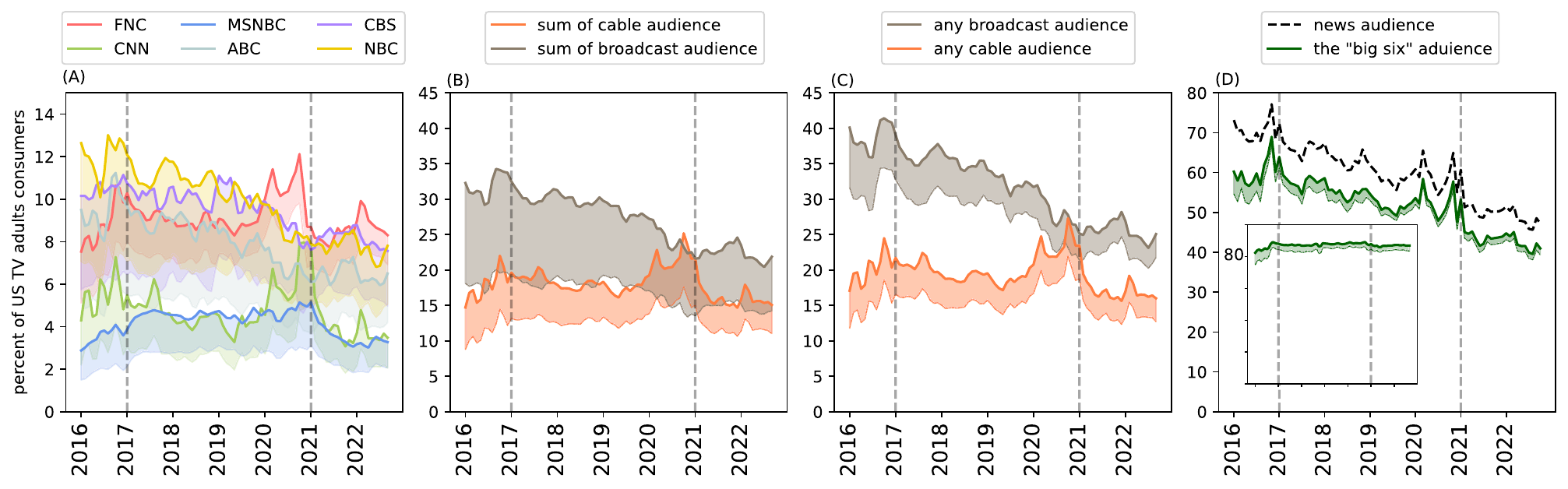}
\vspace{3mm}
\caption{(A) the share of Americans whose TV news consumption is primarily devoted to one station; (B) the sum of the shares of the three broadcast stations and the three cable stations; (C) the share of Americans whose TV news consumption is primarily devoted to either cable (pooling FNC, MSNBC, CNN) or broadcast (pooling ABC, CBS, NBC); (D)  the total share of the US population that consumes at least 30 minutes of news from any source and the share of Americans whose TV news consumption is primarily devoted any combination of the ``big six'' (inset shows the percentage of the total that is accounted for by the big six). Across all panels, solid lines represent the proportion of the U.S. adult population that consumes more than 30 minutes of television news in a given month with 50\%+ of it in that station or collection of stations, while the dotted lines represent 75\%+. Vertical dashed lines reflect 2016 and 2020 presidential elections.
}
\label{fig:segregation}
\end{figure*}

The remainder of Fig.~\ref{fig:polarization_over_time} shows that these three trends--- low and not increasing polarization for broadcast networks, higher and increasing polarization between cable and broadcast, and divergence between FNC and the other two cable networks--also apply separately to hard news (panels D-F) and talk/opinion shows (panels G-I). In addition, we make two further observations. First, the increase in polarization between cable and broadcast news that is apparent in Fig.~\ref{fig:polarization_over_time}B is driven more by increased polarization among hard news programming (Fig.~\ref{fig:polarization_over_time}E), than by opinion/talk shows (Fig.~\ref{fig:polarization_over_time}H). Indeed, while hard news and talk/opinion shows exhibited roughly the same polarization in 2013, hard news was noticeably more polarized than talk/opinion shows by 2023, largely on account of a sharp jump that occurred in 2016. Second, within cable networks, a somewhat opposite result applies: in this case talk/opinion shows contribute more to the overall increase---again, driven by a sharp increase in 2016---and are more polarized overall than hard news programs.

\paragraph*{Topic-level polarization} Moving beyond these station-level trends over time, Fig.~\ref{fig:polarization_top_level} shows average polarization by topic, program type, and year (3 time periods of before 2016 [2012], 2016-2020 [2016] and after 2016 [2020]). 
Fig.~\ref{fig:polarization_top_level}A shows topic-level polarization for hard news/coverage of live events for FNC (red), CNN (green), and MSNBC (blue) versus broadcast news in those topics (as before, we compare individual cable news networks with an aggregation of the broadcast  networks' news content) and Fig.~\ref{fig:polarization_top_level}B shows the same for talk/opinion shows. 
In both cases, Fig.~\ref{fig:polarization_top_level} shows considerable variation across topics. Starting with hard news (Fig.~\ref{fig:polarization_top_level}A), topics such as government and the economy start at relatively low values of polarization and increase by relatively small amounts over the time period, whereas other topics such as crime and climate change, jump to much higher levels in 2016 and 2020. To illustrate the difference, FNC's least polarized discussions of climate change, e.g., ``...\textit{Vice president Al gore on `Meet the Press' not mincing his words about the severity of the global climate crisis and America's inaction}...'' (0.48) are indistinguishable from broadcast news, whereas it's most polarized discussions, e.g., ``...\textit{The real cause of Islamic terror is global warming. Can you agree with that? No, I can't, but that's apparently what the president authentically believes}...'' (0.70) are highly distinctive.  
For talk/opinion shows, meanwhile (Fig.~\ref{fig:polarization_top_level}B), the differences across topics are even more stark. On the one hand, the low polarization topics look very similar to their hard news counterparts, starting off low and not increasing by much. On the other hand, the high polarization topics start off more polarized and increase by even larger amounts (e.g.,  Tucker Carlson on climate change: \textit``{That screeching moron Ocasio-Cortez isn't telling China to give up coal and oil and natural gas and nuclear power within a decade. The left isn't doing any of that because the left loves the Chinese government. It's their model for governing}'' [0.75]). In other words, while the overall increase in polarization is driven more by the increase in hard news (see Fig.~\ref{fig:polarization_over_time}E), the most polarizing coverage still appears on talk/opinion shows where it is concentrated on certain topics.

Shedding more light on this last point, Fig.~\ref{fig:program_polarization} shows polarization of the top five most popular programs identified by Kim et al.~\cite{kim_lelkes_mccrain} for FNC (Fig.~\ref{fig:program_polarization}A), CNN (Fig.~\ref{fig:program_polarization}B), and MSNBC (Fig.~\ref{fig:program_polarization}C) broken out by topic (again versus all broadcast news in the topic). Overall FNC has the highest number of highly polarizing shows---notably Hannity, The O'Reilly Factor, Tucker Carlson Tonight, and the Laura Ingraham Angle---however, a handful of MSNBC shows, led by Deadline: White House (with Nicolle Wallace), are equally polarized. Finally, while CNN's leading talk/opinion shows are overall less polarized than either FNC or MSNBC, Cuomo Prime Time is among the most polarized shows across the three networks. Finally, we note that FNC popular shows are more polarized on topics such as immigration and taxes, while CNN and MSNBC popular shows are most polarized on elections, Russia, and congress. For example, Hannity features highly polarized language when discussing immigration (``...\textit{some democrats even proposed a measure that would force the release of thousands of criminal illegal aliens, including dangerous felons convicted of rape, sex trafficking, violent assault, and even murder into our country. can you believe it?}...'' [0.62]) whereas Deadline: White House uses the most polarized  language around elections (``...\textit{the demise of democracy is happening in full view, predicated on a lie, designed to make it harder for people to vote and to politicize the people who count the votes}...'' [0.63].)

\section*{TV news consumption}
The results of the previous two sections provide unequivocal support for previous claims that cable news stations have been a polarizing influence on the news ecosystem. Not only have cable networks as a whole diverged from their broadcast peers, both in terms of their choice of topics and the language they use to talk about them, but the cable networks have also diverged themselves along partisan lines, with FNC on the right and the increasingly similar MSNBC and CNN on the left. At the same time, however, our results also show that the ``big three'' broadcast networks have remained strikingly consistent, both with respect to each other and over time. ABC, NBC, and CBS, that is, tend to cover similar topics and do so using relatively similar language, and these similarities have changed little over the past decade. Thus, to the extent that viewers remain wedded to the broadcast networks shared reality may be preserved in spite of the polarization of the cable networks. Conversely, to the extent that viewers have migrated from broadcast to cable, then indeed shared reality has been diminished. 

Fig.~\ref{fig:segregation} addresses this concern directly, showing the consumption trends for active news consumers, defined as US adults who consume at least 30 minutes of television news per month, as a fraction of the total US adult television audience (i.e. the total number of adults represented by our panel). We normalize by the adult television audience rather than the US adult population in order to avoid confounding by so called ``cord-cutting'' in which households switch from cable or satellite providers to some combination of antennae and internet-based streaming services, in which case they would no longer appear in the panel. 
Fig.~\ref{fig:segregation}A shows the share for each of the stations individually; Fig.~\ref{fig:segregation}B shows the sum of the shares of the three broadcast stations (grey) and the three cable stations (orange); Fig.~\ref{fig:segregation}C shows the share of the population that get the majority of their consumption from any combination of broadcast-only (grey) or cable-only (orange) stations; and Fig.~\ref{fig:segregation}D, shows the total share of the US population that consumes a majority of its news from any combination of the ``big six'' stations that we study here (green) as well as the total population of news consumers (black dashed line), defined as consuming at least 30 minutes of news from any source, where inset shows the percentage of the total that is accounted for by the big six. Fig.~\ref{fig:segregation} reveals three main findings regarding trends in television news consumption. 

First, Fig.~\ref{fig:segregation}A shows that the share of consumption for the three broadcast networks has generally trended down over the seven-year period of our data, whereas the share for the three cable networks has generally trended up. Consistent with prior work, the broadcast networks are roughly comparable to one another in popularity~\cite{pew_news} while FNC is roughly the size of MSNBC and CNN combined~\cite{muise2022quantifying}. Interestingly, all three cable networks peaked in popularity in 2020, dropping after the 2020 US presidential election, where the post-election drop for FNC is especially dramatic. In part, the overall drop likely reflects a general diminishment of interest in political news following what was an especially consequential election, but in part the drop for FNC likely reflects the widely reported audience backlash against the network for its early call of the election for Joe Biden. Interestingly, we see no equivalent drop for the broadcast networks, suggesting that the more partisan style of cable news is a benefit in politically turbulent times and a liability when the news is more ``boring.''

Second, Fig.~\ref{fig:segregation}B and C both confirm that the total audience for broadcast news has declined substantially over the seven-year period while the total audience for cable has increased slightly. In both cases, however, the total broadcast audience is larger than the total cable audience for the entire period with the exception of the 2020 election, during which cable very briefly matched broadcast in popularity. After the election, the demand for cable news dropped dramatically while the decline in broadcast news has abated somewhat; thus, as of 2023 broadcast news continues to outrank cable. These trends are more pronounced in Fig.~\ref{fig:segregation}C, which shows the audience that gets a majority of its news from any combination of broadcast/cable stations, than in Fig.~\ref{fig:segregation}B which shows the simple sums of the individual audiences. The most likely explanation for this difference is that cable news consumers tend to get their news exclusively from one station whereas broadcast consumers sometimes get their news from more than one station, hence the population of majority broadcast consumers is larger than the sum of the majority consumers for the individual networks (i.e. because the sum of their consumption across broadcast stations can be a majority of their consumption even if consumption of any one station is not). In other words, in spite of its decline and in contrast with conventional wisdom \cite[e.g.]{yglesias18}, broadcast news as a whole remains the primary source of news for Americans. This result is all the more notable in light of the far greater amount of time devoted to news by the cable networks: whereas FNN, CNN, and MSNBC all devote essentially the entire 24-hour day to news coverage, news programs only account for between 8 and 11 hours per day for ABC, NBC, and CBS.

Third, Fig.~\ref{fig:segregation}D shows that, as expected, the audience getting a majority of their television news from any combination of our six stations (green) has also dropped over the seven years of our data, by roughly the same amount as the drop in the combined broadcast audience. Interestingly, however, the total audience of news consumers (dashed line) dropped even more dramatically, by roughly twenty percentage points (from 73\% to 47\%). Put the other way around, the fraction of the television viewing population that consumes less than 30 minutes of news per month roughly doubled, from 27\% to 53\%. Noting that 30 min per month is, on average, one minute per day, this result suggests a doubling in just seven years of people who do not appear to be consuming any appreciable amount of news on television. Previous work~\cite{allen2020evaluating,muise2022quantifying} comparing television with online news consumption has found that television dominates online by a factor of roughly five to one, and that online has not increased appreciably in recent years. Moreover, while cord cutting has eroded the ``traditional'' audience (i.e. cable and satellite subscribers), we note that this drop in news consumption is taking place within the population of traditional television viewers. For both reasons, this result represents a real diminishment in news consumption among the majority of Americans.

\section*{Discussion}
Our results present a mixed picture of the state of shared reality. 
On the production side, we find that the three major cable news channels are increasingly polarized relative to the broadcast networks, covering different topics and using increasingly different language to talk about them~\cite{joshi16, bump22, sullivan19}. Within cable, we also find that FNC has diverged increasingly from CNN and MSNBC, which have simultaneously converged. Surprisingly, we find that these differences appear to be driven more by the increase in polarization of hard news programming than by talk/opinion shows, although the latter appear to host the most polarizing content. These results reinforce conventional wisdom and scales previous work~\cite{kim_lelkes_mccrain,broockman2022impacts} on the polarizing effects of cable news to the majority of segments on cable across nearly 10 years (while also adding contrast with broadcast news). Watching slanted news can shift viewers' attitudes and behaviors---even for dedicated partisans~\cite{broockman2022impacts}. While partisan news has employed us-versus-them frames in political coverage for decades \cite{levendusky13}, the recent divergence in cable networks' topic selection and framing (see Fig.~\ref{fig:aggregate_topics_selection}) seems likely to amplify its polarizing effects. As partisan bias has become more prevalent in cable news production, one-channel cable news viewers are increasingly exposed to a set of facts, narratives, and arguments that have been curated with a consistent partisan goal and lack counterarguments \cite{levendusky13}, all of which is designed to prime their audiences to approve of the conduct of favored politicians and disapprove of opponents.

At the same time, we also find that the ``big three'' broadcast networks (ABC, CBS, and NBC) are largely interchangeable, covering a similar mix of topics and talking about them in a similarly neutral manner, and that these similarities are stable. Thus, the state of shared reality depends heavily on the consumption side: to the extent that Americans get their news from the ``big three,'' they likely do experience a shared sense of reality, whereas to the extent they segregate themselves by cable channels, they likely do not. From this perspective, our findings are ambiguous. On a somewhat positive note, although we find that the share of the population who get a majority of their news from a cable news network has increased somewhat and the corresponding share of majority broadcast news consumers has dropped, the audience receiving a majority of their news from any combination of broadcast networks remains roughly 50\% larger than the corresponding audience for cable. In this sense, therefore, the relatively neutral and homogeneous broadcast networks continue to provide a haven of shared reality for millions of Americans, albeit not as many as in the past. On a more negative note, however, the shrinking of the audience for broadcast news is substantial and appears to reflect a more general trend away from news consumption of any kind: the population of traditional television viewers who watched less than one minute of news per day on average has more than doubled from about 27\% to more than 53\%. Thus, while it is good news from a shared reality perspective that the relatively polarizing cable networks have not yet come to dominate the relatively neutral broadcast networks, the overall turn away from news in general and broadcast news in particular remains very concerning. 

In closing, we note three important limitations of our analysis that we hope will motivate future work. First, by focusing on conventional television viewers, our analysis fails to account for the behavior of so-called cord-cutters who have replaced cable and satellite subscriptions with some combination of antennae and streaming services, both of which offer access to news programming. Our panel of television viewers is representative of at least 80\% of the US adult population; hence we do not expect the behavior of cord cutters---even if it differs from the behavior of cable and satellite subscribers---to appreciably alter our results. Nonetheless, future work should consider the behavior of this now-substantial and growing segment of the population as an object of interest in itself. Second, we have only studied English-language television programming, a restriction that omits several popular Spanish-only networks such as Univision and Telemundo, which collectively account for roughly 5\% of the overall audience for television news. As with cord cutters, this population is likely too small to affect our overall results; however, the similarities and differences of Spanish vs. English language news would also be an interesting avenue for future research. Finally, the apparent disengagement from news even among the conventional television audience raises an unresolved puzzle: if they are not getting their news from television and are not compensating with identifiable online news sources, then from where do they get their information? If they are coming to rely on lower quality of more partisan sources than are available on broadcast television then the effects on shared reality could be substantial, a concern that is amplified by recent evidence suggesting that viewers departing the broadcast news ecosystem are more politically moderate \cite{broockman2023selective}. Alternatively, if they are genuinely disengaging with news altogether, the ongoing exodus may erode political interest, knowledge, and engagement in a large swath of the U.S. electorate \cite{prior2013}, with unknown consequences. A final area of future work, therefore, is to better understand this trend away from traditional news and what is replacing it.

\section*{Materials and Methods}

\subsection*{Segment classification}

As a first step, we remove all the advertisements in our entire transcript data set. 
Since each episode of a news program may cover multiple topics, we split each ad-free episode transcript into ``segments'' that can contain up to 150 words, resulting in a total of $13,446,736$ segments across all six stations. The next step is identifying topics discussed in each segment. While topic models can be very effective at uncovering topics within large collections of documents, scholars have expressed concern about measuring concepts using topic models without assessing their accuracy~\cite{hoyle2021automated,ying2022topics}. To rigorously quantify how major TV stations cover important topics of the day, we propose a novel two-layered human-in-the-loop multilabel classification model, where a segment can belong to none, one, or multiple of the predefined labels. 

Because we classify segments across a wide range of topical dimensions, traditional approaches to supervised classification would be infeasible. Training and validating a multilabel model or a series of binary classifiers would also require extensive annotated data, which would be inefficient and costly: (i)~it would require annotators to select a topic from a large list of candidates, (ii)~considering that class imbalance varies across topics (and is likely to be high for the majority), it would be necessary to sample approximately 10,000 segments for a topic covered for less than 5 percent of the total news airtime in order to collect 500 positive results. Instead, we propose a two-layer human-in-the-loop approach whereby, in the first layer, a weakly-supervised classifier with high recall is utilized to narrow the search space for segments belonging to each topic. The objective of this layer is to identify samples that are highly likely to be positive for each class on the basis of as little information as one word per topic. The first layer produces a set of candidate segments for each topic that are high in recall but low in precision. By removing irrelevant segments from consideration for each topic, the first layer greatly increases the true positive ratio for each topic, improving class imbalance. As a result, supervised classification performed in the second layer is more efficient. Within each topic, we select a random set of segments for annotation from the reduced set, then ask an annotator to select from a limited number of three topics, including the one that has the highest probability. The annotator can indicate that a segment should be classified as none or one of the provided topics. With the human-labeled segments as our ground truth, we develop supervised models in the second layer to refine low-precision classifications from the first layer.

\paragraph{Weakly-supervised layer}
Keyword matching is one of the common methods used to classify documents into a set of predefined topics. It is, however, limited by low recall---(i) there are several (or many) keywords that may be representative of a topic, (ii) a topic may be discussed without the use of common keywords---and low precision---(i) the use of a topic-relevant word does not necessarily indicate it is the topic of the document; (ii) the same word can have a variety of meanings (for example, lead poisoning vs. lead officer). To overcome these shortcomings, we develop an ``expanded dictionary'' for each topic containing words semantically similar to its corresponding label word, as proposed by Meng et al.~\cite{meng2020text}. Then, each segment is assigned to a topic if it contains words semantically similar to those in the topic's expanded dictionary. This approach assumes that, inherently, words with similar meanings are interchangeable. Thus, to create an expanded dictionary for each topic, we identify a suite of words that are semantically close to the topic label. To populate each expanded dictionary---by finding the most likely words that would replace each topic label in context---we use a pre-trained masked language model (MLM)~\cite{kenton2019bert}. For each topic, we successively mask each occurrence of the label word in the corpus and then feed the resulting contextualized embedding vectors from the BERT encoder to the MLM head. The output is a probability distribution over the entire vocabulary $V$, which indicates the likelihood of each word $w$ appearing at the given position where the topic label word was masked. Our system loops over each segment and stores the top 50 probable replacements for each label occurrence. Next, for each topic, we rank the candidate words by the number of times they appear in a top-50 most probable replacements list and create a ``class vocabulary'' set for each topic. To do so, we keep the top 100 most frequent replacement words. After creating such a vocabulary for each topic, the words within each set are manually reviewed by authors, and the ones that are too general or irrelevant to the label of interest are removed. The resulting list for each topic comprises our expanded keywords.

By utilizing the developed class vocabulary, each word in the corpus is examined within its context. Each document $d$ is subjected to the following process: (1) mask each word $w$ in document $d$, (2) for each masked word $w$, predict the top 50 words that can replace it, (3) determine the overlap between the predicted set for word $w$ and class vocabulary for label $z$, and (4) assign the corresponding topic to document $d$ when the overlap exceeds a predefined threshold for at least one word in the document~\cite{meng2020text}. Each segment passes through this layer and obtains a label per topic that indicates that it either belongs or does not belong to the corresponding class label. Consequently, each segment may be associated with multiple topics. It is important to note that a high threshold creates a more stringent limit for topic assignment, which may result in greater precision at the risk of losing true positives, and vice versa for a low threshold. Because the first layer optimizes for recall, we set a lenient threshold of 20\% overlap, which results in a greater number of false positives. This first layer greatly reduces the search space for each topic $z$ and improves its class imbalance (from $\mathcal{D}$ to $\mathcal{D}^{\text{weak-sup}}_z$), resulting in a more efficient second-stage annotation process.

\paragraph{Topic annotation}
We proceed by selecting 50 random segments from each topic-station, for a total of 300 segments per topic $z$ from the corresponding set $\mathcal{D}^{\text{weak-sup}}_z$. Using the Amazon Mechanical Turk crowdsourcing service, four annotators label each segment---where we break ties with more annotations---summing to a total of almost 30 thousand annotations. A majority vote is required for the assignment of a topic to a segment. The labeled data is subsequently used as ground truth to train and validate supervised classifiers in the second layer. We also evaluate the precision of first-layer predictions, which is a weakly-supervised model where we only feed the class label names into the model. 

\paragraph{Supervised classification layer}
The inputs for the second layer are set $\mathcal{D}^{\text{weak-sup}}_{(\text{station},z)}$ and the ground truth obtained from human annotators. For each station-topic pair, we perform 5-fold cross-validation in the training phase and choose the model with the highest F1 score. Then, passing all the segments in set $\mathcal{D}^{\text{weak-sup}}_{(\text{station},z)}$ through its corresponding trained model, we create a set containing segments with the positive class label as $\mathcal{D}^{\text{sup}}_{(\text{station},z)}$. We achieve an average precision of $78.45\% \pm 13.65$ over $24\times6$ models. 
All the results in the main text are based on the output of the second layer ($\mathcal{D}^{\text{sup}}_{(\text{station},z)}$), which achieves far higher precision than the outputs from the first, weak-supervised layer alone.

\subsection*{Measuring polarization}
\label{polarization}
Following Refs. \cite{gentzkow2019measuring} and \cite{demszky2019analyzing}, which study partisanship within congressional speeches and social media polarization respectively, we define polarization within our study as the probability that an observer with a neutral prior will assign a segment to its true station after observing a unigram or bigram from that segment. When considering a given representation $x$ of segment $d$, we define the degree of polarization as the divergence between $x^{target} (d)$ and $x^{source} (d)$. When these vectors are close in the representation space, this indicates that the source and target speak in a similar manner, and we refer to this as polarization is low. When the language used by the source and target are more distinct, the vectors will be more distant, indicating higher polarization. To measure the partisanship between two stations, we use the leave-out estimator from Gentzkow et al.~\cite{gentzkow2019measuring}, defined as: 
\begin{equation*}
\hat{\pi}^{LO} = \frac{1}{2}\sum_{i\in source} \hat{q}_i . \hat{\rho_{-i}} + \frac{1}{2}\sum_{i\in target} \hat{q}_i . (1 - \hat{\rho_{-i}})
\label{polarization_equation}
\end{equation*}

For segment representation, we use the vector of unigrams and bigrams (phrases) frequencies, where $\hat{q}_i = \textbf{c}_{i}/m_i$, $\textbf{c}_{i}$ is the vector of phrase counts and $m_i=\sum c_{ij}$ is the total number of phrases in segment $i$. Note that equation \ref{polarization_equation} is symmetrical and there is no order to source and target stations. Furthermore, we remove possible confounders, e.g., for a given station we remove program names, hosts' names, and mentions of the station name itself. Let $\hat{q}^{G} = \sum\limits_{i\in G} \hat{\textbf{c}}_i / \sum\limits_{i\in G} \hat{m}_i$ is the empirical term frequencies for station $G $ and
$\hat{\rho} = \hat{q}^{source} \oslash (\hat{q}^{source} + \hat{q}^{target})$, where $\oslash$ denotes element-wise division. Then, $\hat{\rho_{-i}}$ is the analogue of $\hat{\rho}$ computed from the empirical frequencies where segment $i$ is excluded. The segment-level estimate of the partisan score for segment $i$ is also defined as the dot product $\hat{q}_i \cdot \hat{\rho}_{-i}$,~\cite{demszky2019analyzing}.

\begin{acknowledgements}
\textbf{Acknowledgments}: We are grateful to TVEyes for access to their television transcript archive and to the Nielsen Company for access to their television panel data. In addition we are grateful to B. Sissenich, S. Sherman, H. Baberwal, and E. Grimaldi of the Nielsen Company for ongoing support, to Amir Ghasemian for helpful conversations, and to Maya Jamdar, Lindsey Perlman, Anika Prakash, Pia Singh, Sina Shaikh, Kaelin Suh, and Kira Wang for research assistance.
Finally, H.H., and D.J.W. are grateful for the financial support provided by Richard Jay Mack and the Carnegie Corporation of New York (Grant G-F-20-57741). 
\end{acknowledgements}


 

%

\end{document}